\documentclass[letterpaper,english,aps,prb,floatfix,twocolumn,showpacs,amsfonts,amssymb]{revtex4}

\usepackage[T1]{fontenc}
\usepackage[latin1]{inputenc}
\usepackage{graphicx}
\usepackage{amssymb}
\usepackage{overpic}

\usepackage{babel}

\newcommand{\beq}{\begin{equation}}
\newcommand{\eeq}{\end{equation}}
\newcommand{\bea}{\begin{eqnarray}}
\newcommand{\eea}{\end{eqnarray}}

\newcommand{\kb}{{\bf k}}
\newcommand{\qb}{{\bf q}}

\begin{document}
 
\title{Renormalization Group Approach to Strong-Coupled Superconductors}

\author{S.-W.~Tsai$^1$, A.~H.~Castro~Neto$^1$, R.~Shankar$^2$, 
and D.~K.~Campbell$^1$}

\affiliation
{$^1$ Department of Physics, Boston University, Boston, MA 02215\\
$^2$ Sloane Laboratory of Physics, Yale University, New Haven, CN 06520}

\date{\today}

\begin{abstract}
We develop an asymptotically exact renormalization group (RG) approach
that treats electron-electron and electron-phonon interactions
on an equal footing. The approach allows an unbiased study of the instabilities
of Fermi liquids without the assumption of a broken symmetry. We apply our
method to the problem of strongly coupled superconductors and find the
temperature $T^*$ below which the high-temperature Fermi liquid state
becomes unstable towards Cooper pairing. We show that $T^*$
is the same as the critical temperature $T_c$ obtained in Eliashberg's
strong coupling theory starting from the low-temperature superconducting
phase. 
A $1/N$-expansion shows that the method is asymptotically exact and 
Migdal's theorem follows as a consequence. 
Finally, our results lead to a novel way
to  calculate numerically, from microscopic parameters, the transition
temperature of superconductors.
\end{abstract}


\maketitle

There is a renewed interest in understanding the interplay between
electron-electron and electron-phonon interactions in strongly
correlated electronic systems
\cite{shen,organics,mgb2,millis,c60}. While the experimental data
indicate the important role played by both electron-electron and
electron-phonon interactions, theoretical progress has been
limited due to the complexity associated with treating these two
interactions on an equal footing. Most approaches rely on mean-field
treatments where a ``favorite'' order parameter and broken
symmetry are introduced ``by hand''. It is highly 
desirable if instead one could predict the leading instability starting
from the geometry of the Fermi surface, the strength of the
couplings, and the bare energy scales of the problem, such as the
Fermi energy, $E_F$, and the Debye frequency, $\omega_D$.

The renormalization group (RG) provides such an approach \cite{shankar}, one  that has
been successful in explaining the stability and instabilities of
Landau Fermi liquids in more than one dimension. 
Let us recall
the two-stage procedure advocated there. Firstly, given a
microscopic theory defined in all of momentum space one integrates
out all modes except those within an energy cut-off $\Lambda$ of $E_F$. In
$d=2$, which is our focus in this work (the $d=3$ case can
be treated in analogous way \cite{next}), the remaining phase 
space has the form of an annulus of radius $k_F$ (the Fermi momentum) 
and width $2 \Lambda/v_F$ (where $v_F$ is the Fermi velocity). 
In the beginning of this process of integrating out high energy modes, the 
corrections are treated perturbatively in the strength of the interactions.
In the second stage, provided that mode elimination have reached an energy
cut-off $\Lambda \ll E_F$, a $1/N$ expansion emerges, 
with $N\simeq E_F/\Lambda$. 
More precisely, imagine dividing the
annulus into $N$ patches of size of order $(\Lambda/v_F)^2$. The momentum
of each fermion ${\bf k}$ is a sum of a "large" part (order
$k_F$) centered on a patch labeled by a patch index $i=1,..., N$
and a "small" momentum (order $\Lambda/v_F$) within the patch
\cite{shankar}. It can then be verified that in all Feynman
diagrams of this cut-off theory the patch index plays the role of
a conserved isospin index exactly as in a theory with $N$
fermionic species. The electron-electron interaction terms,
written in this notation, come with a pre-factor of $1/N$
($\sim \Lambda /E_F$), and the RG corrections can be organized in terms of 
powers of $1/N$. Summing up the series of dominant corrections in order
$1/N$ becomes asymptotically exact since $N \to \infty$ as the RG
procedure decreases the cut-off $\Lambda \to 0$.  This $1/N$-expansion
has been developed in ref.[\onlinecite{shankar}] and has been used to 
show that, provided one can start the RG flow at $\Lambda_0 \ll E_F$ (large
$N$), Landau's Fermi liquid theory emerges
as the asymptotically exact theory for repulsive fermions with a generic
(non-nested and with no singularities) Fermi surface. 
In this way the RG
expansion goes beyond perturbation theory and the end values of the couplings
can be large. We make use of this expansion here and extend it for the case
when phonons are present.   

Electron-phonon interactions in the Wilson-like RG of ref.
[\onlinecite{shankar}], wherein momenta are rescaled to attain a fixed
point, posed the following problem \cite{polchinski}: 
electronic momentum scales
differently parallel and perpendicular to the Fermi surface while
phonon momentum scales isotropically.   We circumvent this problem
by  using the quantum field theory version of the RG in which
the cut-off dependence of couplings that preserves the physical
quantities defines the flow, with no rescaling of momenta or
frequencies. 

The paper is organized as follows: 
The RG approach for interacting fermions that are 
also coupled to phonons is described in Sec. \ref{sec:1}.
In Sec. \ref{sec:2} we illustrate our method for a circular Fermi surface and
an analytical solution is obtained.  The RG
indicates an instability in the Cooper channel when the
electron-phonon coupling is strong enough to overcome the
effective repulsive electron-electron interactions. In Sec. \ref{sec:3} we
present the finite temperature formalism. We obtain the
temperature $T^*$ at which the high-temperature Fermi liquid state
becomes unstable in the Cooper channel. We demonstrate that $T^*$
is the {\it same} as the superconducting temperature $T_c$ that is
obtained from the Eliashberg theory of strongly coupled
superconductor, which approaches  the transition from the ordered
phase \cite{eliashberg,carbotte}. 
This is an alternative 
derivation of Eliashberg's equations starting from a Fermi
liquid state. Furthermore, by extending the $1/N$ analysis to the problem
with phonons (Sec. \ref{sec:4}), we show that, as in the case of
the Landau's Fermi liquid theory for electron-electron
interactions, the Eliashberg theory is the exact low energy
effective theory obtained by using RG. Migdal's theorem is also derived from
the $1/N$ analysis. Sec. \ref{sec:5} contains further discussions and
conclusion. 

\section{RG for interacting electrons coupled to phonons}
\label{sec:1}

We work in the path-integral representation and consider the
general action 
\begin{eqnarray}
S(\psi,\phi) = S_e(\psi) + S_{ph}(\phi) +
S_{e-ph}(\psi,\phi) + S_{e-e}(\psi)
\end{eqnarray} 
where 
\begin{eqnarray}
S_e = \int_{\omega \kb}
\psi^{\dagger\sigma}_k (i\omega - \epsilon_{\kb}) \psi_{k \sigma}
\end{eqnarray}
is the free electron action and
\begin{eqnarray}
S_{ph} = \int_{\Omega \qb} \phi^{\dagger}_{q} (i\Omega - w_{\qb})
\phi_{q}
\end{eqnarray}
is the free phonon action ($k = \{\omega, \kb\}$ and $q = \{\Omega, \qb\}$,
where $\omega,\Omega$ are fermionic and bosonic Matsubara frequencies,
respectively, and $\kb,\qb$ are the momenta). 
The interaction terms are given by 
\begin{eqnarray}
S_{e-ph} =
\int_{\omega \kb} \int_{\Omega \qb} g(q) \psi^{\dagger \sigma}_{k+q}
\psi_{k \sigma}(\phi_q + \phi^{\dagger}_{-q})
\end{eqnarray}
and
\begin{eqnarray}
S_{e-e} = \frac{1}{2}
\prod_{i=1}^3 \int_{\omega_i \kb_i} u(k_4,k_3,k_2,k_1)
\psi^{\dagger \sigma}_{k_4} \psi_{k_2 \sigma} \psi^{\dagger
  \sigma^{\prime}}_{k_3} \psi_{k_1 \sigma^{\prime}} 
\end{eqnarray}
where $k_4=k_1+k_2-k_3$. (We use units such that $\hbar = 1 =
k_B$.) The above action defines the input physics at a cut-off
$\Lambda_0$ such that $\omega_D, g, u \ll \Lambda_0 \ll E_F$ (thus $N\simeq
E_F/\Lambda_0 \gg 1$).

\begin{figure}
\includegraphics[bb=180bp 320bp 418bp 466bp,clip,scale=1]{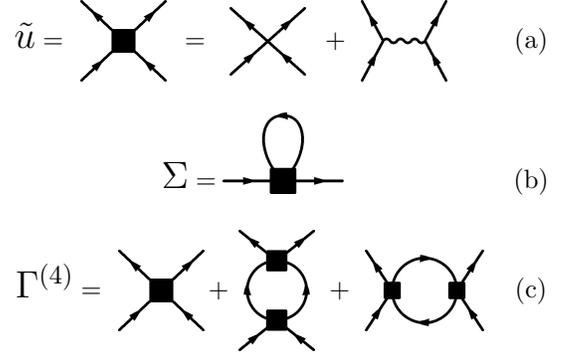}
\caption{(a) The retarded interaction $\tilde{u}$; (b) The self-energy
  correction; (c) The interaction vertex.}
\label{fig:u}
\end{figure}

The phonons, being described by a gaussian action, can be integrated 
out exactly leading to an electron-electron problem with retarded
interactions
\begin{eqnarray}
\tilde{u}(k_4,k_3,k_2,k_1) &=&
u(k_4,k_3,k_2,k_1)
\nonumber
\\
&-& 2 g(k_1,k_3) g(k_2,k_4) D(k_1-k_3) \, ,
\label{retarded}
\end{eqnarray}
where
\begin{eqnarray}
D(q) = \frac{\omega_{{\bf q}}}{\omega^2+\omega_{{\bf q}}^2}
\end{eqnarray}
is the phonon propagator. Here we are considering fermions with spin and it
is sufficient to focus on interactions involving two fermions with opposite
spins\cite{Zanchi}. For spinless fermions the initial condition must be
antisymmetrized with respect to 1$\leftrightarrow$2 and
3$\leftrightarrow$4. The Feynman diagram associated with the 
initial vertex $\tilde{u}$ is shown in Fig.\ref{fig:u}(a). 
The vertex is a function of the momenta and frequencies of the in-coming and
out-going electrons, with momentum and frequency conservation. The dependence
on the magnitude of the momenta is irrelevant and all ${\bf k}$'s are taken
to be on the Fermi surface. 
For a generic non-nested Fermi surface, with no singularities and with
time-reversal symmetry, only two types of scattering exist \cite{shankar}:
forward scattering with $k_1=k_3,\ k_2=k_4$  (these
evolve into Landau parameters) and the scattering
in the Cooper channel with $k_1=-k_2,\ k_3=-k_4$. The box vertex in
the diagrams in Fig.\ref{fig:u} can represent either one of these 
scattering processes.
The forward scattering channel (which does not flow under the RG
as in the case of pure electron-electron interactions 
\cite{shankar}) contributes to the electron self-energy 
$\Sigma({\bf k},\omega)$ as shown in Fig.\ref{fig:u}(b). 
We can write 
\begin{eqnarray}
\Sigma(\omega,{\bf k}) = \Sigma_0 + i (1-Z(\omega,{\bf k})) \omega
\end{eqnarray} 
with two types of contributions: a shift in the chemical potential ($\delta \mu
\propto \Sigma_0$) and wave-function renormalization, $Z(\omega,{\bf k})$.
The shift in the chemical potential can be reabsorbed in the
theory by assuming a fixed number of electrons \cite{shankar}. The
wave-function renormalization, $Z(\omega,{\bf k})$, is of special interest
in this problem.

The starting point of our RG is the interaction vertex
$\Gamma^{(4)}[\tilde{u}]$ in the Cooper channel, as
shown in Fig.\ref{fig:u}(c):
\bea
\Gamma^{(4)}[\tilde{u}(-k_3,k_3,-k_1,k_1)] = \tilde{u}(-k_3,k_3,-k_1,k_1)  
\hspace{1cm}
\nonumber \\
- \int_{\omega \kb} \frac{\tilde{u}(-k,k,-k_1,k_1)
    \tilde{u}(-k_3,k_3,-k,k)}{
(i \omega - \epsilon_{\kb} - \Sigma(\omega,\kb))( - i \omega - 
\epsilon_{\kb} - \Sigma(-\omega,\kb))}   \, ,
\label{rgcomplete} 
\eea 
where the momentum integral is such that
all internal energies lie between $0$ and $\Lambda$. 

\section{Exact analytical solution for a circular Fermi surface}
\label{sec:2}

In order to 
establish the method and proceed with simple analytical
manipulations, we focus on an isotropic Fermi surface with no singular
regions, and Einstein
phonons with frequency $\omega_E$ (the generalization for an
arbitrary phonon spectrum is straightforward \cite{next}). Since,
$E_F \! \! \gg \! \! \omega_D,g,u$ we ignore  radial excursions away from the
Fermi surface in the coupling constants. In this case the external
electron momentum can be put on $k_F$ and
$\tilde{u}(-k_3,k_3,-k_1,k_1)$ depends on the angles only via the
difference $\theta_1 \! - \! \theta_3$. Let us define
\begin{eqnarray}
\tilde{v}(\omega_1,\omega_3) \! = \! N(0) \int \frac{d\theta_1}{2\pi}
\int \frac{d\theta_3}{2\pi} \tilde{u}(-k_3,k_3,-k_1,k_1)\
\nonumber
\end{eqnarray}
where $N(0)$ is the Fermi surface density of states. 

The RG equations are obtained from (\ref{rgcomplete}) by imposing
the condition of cut-off independence, namely, 
\begin{eqnarray}
\frac{d \Gamma^{(4)}}{d \ell} = 0
\end{eqnarray} 
where $\ell = \ln(\Lambda_0/\Lambda)$ is the RG scale.
From (\ref{rgcomplete}) we obtain:
\bea
\frac{d}{d\ell} \tilde{v}(\omega_1,\omega_3,\ell) = -
\int_{-\infty}^{\infty}
\frac{d\omega}{\pi} \frac{\Lambda_{\ell} \, \, \tilde{v}(\omega_1,\omega,\ell)
  \tilde{v}(\omega,\omega_3,\ell)}{
\Lambda_{\ell}^2+ Z^2_{\ell}(\omega) \omega^2 }
 \, ,
\label{eq:func} \eea 
with  the initial RG condition
\begin{eqnarray}
\tilde{v}(\omega_1,\omega_3,\ell=0) = u_0 -  \lambda \omega_E
D(\omega_1-\omega_3)
\nonumber
\end{eqnarray} 
where $u_0$ is the bare electron-electron
interaction  \cite{screening} and
\begin{eqnarray}
\lambda =  \frac{2 N(0) g^2}{\omega_E}
\end{eqnarray}
is the electron-phonon coupling constant. Notice that
this is a  {\it functional} equation, because retardation
introduced by the phonons  leads to a mixing of the couplings at
low and high frequencies. 

The RG equation for $Z_{\ell}(\omega)$, which for a
circular Fermi surface does not depend on the momentum direction,
can be likewise derived and formally integrated
to give:
\begin{eqnarray}
Z_{\ell}(\omega) = 1 + \frac{\lambda}{\pi \omega}
\int_{\omega'}\int_{\Lambda_{\ell}}^{\infty}
\frac{Z_{\Lambda} \omega' D(\omega-\omega')}{Z^2_{\Lambda}(\omega') \omega^{'2}+\Lambda^2} \, .
\label{big_Z}
\end{eqnarray}

The large-$N$ considerations of ref. [\onlinecite{shankar}] are fully
applicable here since we are dealing with non-singular 
interactions (albeit retarded). Thus the one-loop flows
derived above are exact as $N\to \infty$. More detailed 
analysis is presented in Sec. \ref{sec:4}.

Notice that
(\ref{eq:func}) can be seen as a matrix problem:
\begin{eqnarray}
\frac{d {\bf U}}{d \ell}  = - {\bf U} \cdot {\bf M} \cdot {\bf U}
\label{eq:matrix}
\end{eqnarray}
where 
\begin{eqnarray}
U_{ij}(\ell) &=& \tilde{v}(\omega_i,\omega_j,\ell) \, ,
\nonumber
\end{eqnarray}
and
\begin{eqnarray}
M_{ij}(\ell) &=& \frac{\Lambda_{\ell} \delta_{ij}}{\pi(\Lambda_{\ell}^2+
Z^2_{\ell}(\omega_i) \omega_i^2)} \, .
\nonumber
\end{eqnarray} The solution of (\ref{eq:matrix})
is: 
\begin{eqnarray}
{\bf U}(\ell) &=& \left[1+{\bf U}(0) \cdot {\bf P}(\ell)\right]^{-1} {\bf
  U}(0) \, ,
\nonumber
\end{eqnarray}
with ${\bf P}(\ell)$ defined as
\begin{eqnarray}
{\bf P}(\ell) &=& \int_0^{\ell} d\ell' {\bf M}(\ell')\, .
\nonumber
\end{eqnarray} 
Therefore, the condition for the instability of the RG at a certain
scale $\ell = \ell_c$ is given by:
\begin{eqnarray}
\det\left[1+{\bf U}(0) \cdot {\bf P}(\ell_c)\right] = 0 \, .
\label{eq:det}
\end{eqnarray}

Equivalently, we could search for the eigenvector
${\bf f}$ of the matrix ${\bf U}^{-1}(\ell_c)$ with zero
eigenvalue: 
\begin{eqnarray}
\left[1+{\bf U}(0) \cdot {\bf
    P}(\ell_c)\right] \cdot {\bf f} =0 \, ,
\nonumber
\end{eqnarray}
that is,
\begin{eqnarray}
f(\omega) = - \frac{1}{\pi} \int_{\omega'} \int_{\Lambda_c}^{\infty}
\frac{[u_0 - \lambda \omega_E D(\omega-\omega')]}{Z^2_{\Lambda}(\omega') \,
  \omega^{'2}+\Lambda^2} f(\omega') \, ,
\label{exact_1}
\end{eqnarray}
which is an integral equation for $f(\omega)$.

For a given value of input parameters ($u_0, \lambda, \omega_E$,$\Lambda_0$)
the set of equations (\ref{exact_1}) and (\ref{big_Z}) can be solved for a
critical  
cut-off energy scale,
$\Lambda_c$, at which the running couplings diverge and the
Fermi liquid description breaks down. 

\section{Finite temperature formalism and derivation of Eliashberg's
  equations}  
\label{sec:3}

The $T=0$ formalism presented in the previous section 
can be readily extended to $T>0$, providing us with a
more experimentally accessible quantity namely, a critical temperature. 
We seek the temperature $T^*$ below which the Fermi liquid description 
ceases to exist as one scales towards the Fermi surface, that is, as
$\Lambda_c \to 0$. In this case we replace the integrals in (\ref{big_Z})
and (\ref{exact_1}) by Matsubara sums and extend the integrals in
$\Lambda$ from $0$ to $\infty$ to obtain
\begin{eqnarray}
Z(\omega_n) \phi(\omega_n)\!=\!-\!\pi T^*\!\sum_{m}
[u_0\!-\!\lambda \omega_E D(\omega_n\!-\!\omega_m)]\frac{\phi(\omega_m)}{|\omega_m|},
\label{matsubara_1}
\end{eqnarray}
where we have defined $\phi(\omega_n) = f(\omega_n)/Z(\omega_n)$
and the Matsubara frequency $\omega_n = \pi T^* (2 n +1)$, where $n$ is an
integer.
Moreover, from (\ref{big_Z}) we find:
\begin{eqnarray}
Z(\omega_n)  =1
+ \lambda \omega_E \frac{\pi T^*}{\omega_n} \sum_{m} {\rm sgn}(\omega_m)
D(\omega_n-\omega_m) \, .
\label{exact_Z_matsubara}
\end{eqnarray}
The solution of (\ref{matsubara_1}) and (\ref{exact_Z_matsubara})
gives the value of $T^*$ as a function of the input parameters.

We now relate our approach to Eliashberg's self-consistent
mean-field theory  which assumes a broken symmetry with a
superconducting  order parameter $\Delta(\omega_n)$, in contrast
to ours which starts from the Fermi liquid phase. Remarkably
(\ref{matsubara_1}) and (\ref{exact_Z_matsubara}) coincide with
the Eliashberg equations at $T=T_c$  if we replace
$\phi(\omega_n)$ by $\Delta(\omega_n)$ \cite{carbotte}. This
result is striking since $\phi(\omega_n)$ is not an order
parameter and no symmetry breaking was assumed in our calculation.
By the same token, we can show that $\Lambda_c$ plays the role
of the zero temperature superconducting gap, $\Delta_0$.
Since the RG procedure, approaching the instability from high
temperatures, leads to an instability of the Fermi liquid state at
a temperature $T^*$ that is equal to the critical temperature
$T_c$ produced by the Eliashberg theory. Thus, it is no surprise 
that we can show that in the weak/intermediate coupling regime
($\mu^* \!<\! \lambda \!<\! 1$) we recover the McMillan formula
\cite{mc}: 
\begin{eqnarray}
T^* \approx 1.13 \, 
\omega_E \exp\left\{- \frac{1+\lambda}{\lambda-\mu^*(1+\lambda)}\right\} \, ,
\end{eqnarray}
where 
\begin{eqnarray}
\mu^*\!=\! \frac{u_0}{1 \! + \! u_0 \ln(\Lambda_0/\omega_E)} \, ,
\end{eqnarray}
is the effective electron-electron
interaction at the scale of $\omega_E$ (Anderson-Morel potential) 
and the  Allen-Dynes
expression \cite{allen}: 
\begin{eqnarray}
T^* \approx 0.16 \sqrt{\lambda} \omega_E
\end{eqnarray} 
at strong coupling.

\begin{figure}
\includegraphics[bb=185bp 325bp 428bp 485bp,clip,scale=1]{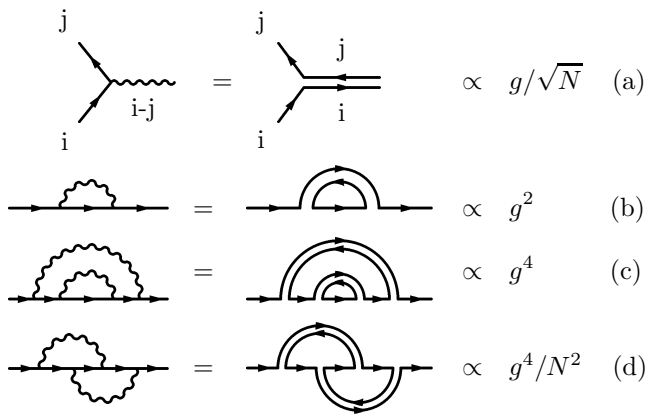}
\caption{(a) The electron-phonon vertex; (b)-(d) are self-energy 
corrections.}
\label{fig:z}
\end{figure}

\section{Expansion in $1/N$ and derivation of Migdal's theorem}
\label{sec:4}

From the $1/N$ analysis of ref.[\onlinecite{shankar}], we know 
that the solution above for the one-loop RG equations 
contains the sum of all dominant corrections in $1/N$. The terms that have
been left out go to zero as $N \to \infty$. 
We have therefore demonstrated that, provided that one can start the RG flow
at $\Lambda_0 \ll E_F$, Eliashberg's theory is the 
asymptotically exact description of the effective low-energy
physics obtained by RG thanks to the small parameter $1/N$.
While this is an important result, the main significance of our methods is
that it can be used to study the competition between charge
density wave (and other instabilities) and superconductivity in the strong
coupling ($\lambda \ll 1$) regime of phonons, something that cannot easily be
achieved in a mean-field approach. This issue arises for example for Fermi
surfaces with regions of nesting. For arbitrary shapes of the Fermi surface
the solution of the RG equations can be obtained numerically by discretizing
the Fermi surfaces into patches\cite{Zanchi,patches}.

Since Eliashberg's theory is based on Migdal's theorem \cite{migdal}, 
which states that electron-phonon vertex corrections to the electron
self-energy vanish as $\omega_D/E_F \to 0$,
it must be that this theorem is built into our approach. We 
show here how this result arises and that the small parameter 
$\omega_D/E_F$ of Migdal's theorem is replaced here by $1/N$.

Let us go back to the problem before we traced the phonons and
discuss the $1/N$ hierarchy. The patch index notation carries very
naturally to phonons. While it is obvious that when a fermion in
patch $i$ scatters to patch $j$, it emits a phonon of momentum
$\kb_j-\kb_i$, we can also go the other way: a phonon of "large" 
momentum can be resolved, up to a two-fold ambiguity, into a
difference ${\bf k}_i-{\bf k}_j$ associated with patches $i$ and
$j$, due to the fact that the electron momenta lie in a thin annulus 
around the Fermi surface. This fact allows us to
describe phonons with the double-index notation employed  by 
t'Hooft for gluons in QCD \cite{hooft}: the phonon line is seen as
made out of two counter-propagating electron lines, as depicted in
Fig.\ref{fig:z}(a). The feature of the large $N$ approach, 
that given a sum or difference of momenta, one can uniquely 
reconstruct the parts as $\Lambda \to 0$ was pointed out in 
ref.[\onlinecite{shankar}]. Since
integrating out the phonons produces a four-fermion interaction of
size $1/N$ it is clear the electron-phonon vertex will be of size
$1/\sqrt{N}$ in our notation. 

Any given diagram made of $n_u$ vertices $u$, $n_g$ vertices $g$
and $n_L$ internal loops is of order $(1/N)^n$ where $n = - n_u -
n_g/2 + n_L$. This way we can organize the diagrammatic expansion 
in powers of $1/N$. The number of loops in each diagram, $n_L$, 
can be easily obtained using Fig.\ref{fig:z}(a).
Consider the problem of the RG for $\Sigma({\bf  k},\omega)$ shown
in Fig.\ref{fig:z}(b)-(d). It is clear from Fig.
\ref{fig:z}(b) that the correction of order $g^2$ should be taken
into account. While $g^2$ comes with a factor $1/N$, there is an
internal closed loop giving an extra factor of $N$ resulting in a
correction of $O(N^0)$ for the self-energy. At the same order in
perturbation expansion in $1/N$, there are diagrams that are made
from a simple repetition of Fig. \ref{fig:z}(b), which are
obviously of order $O(N^0)$ but higher order in $g$. Diagram Fig.
\ref{fig:z}(c) is of order $O(N^0)$, like diagram \ref{fig:z}(b),
and so are all the other ``rainbow'' diagrams that are
automatically  included into the theory.  Thus, by solving the RG
equations for a ``running'' self-energy at one-loop, we in fact
take into account {\it all} corrections of order $O(N^0)$ to {\it
all} loops. The infinite series of diagrams being summed is not
arbitrary and arises naturally from the RG equations, as shown by
our $1/N$ analysis. The final diagram of order $g^4$ is shown in
Fig.\ref{fig:z}(d): this is the famous vertex correction to the
electron self-energy due to electron-phonon interactions studied
by Migdal \cite{migdal}. Notice that diagram \ref{fig:z}(d) does
not contain any internal loops and is of order $O(N^{-2})$ and
thus is vanishingly small as $N \to \infty$. It is easy to show
that these corrections to the self-energy lead to (\ref{big_Z}).
Thus, Migdal's theorem is built into the large $N$ approach
\cite{chubukov}. We can also show that the phonon self-energy 
has only contributions of order $1/N^2$.

\begin{figure}[t]
\centerline{
\includegraphics[
clip,scale=0.42]{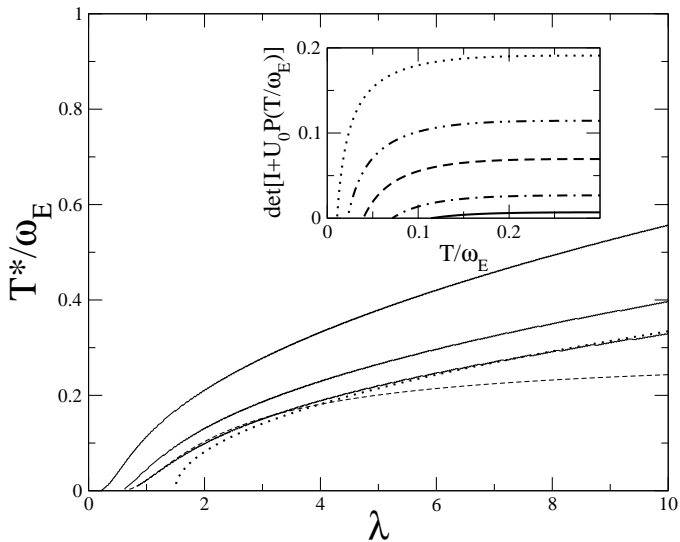}
}
\caption{\label{numerics}The numerical solution of (\ref{eq:det}) 
(continuous line) for $\mu^*=0$, $0.17$, and $0.24$ (from top to bottom)
as a function of $\lambda$. Dashed line: $T^*/\omega_E = 1.1 \,
\exp\{-(1+\lambda)/(\lambda+\mu^*(1+\lambda)\}$ for $\mu^* = 0.24$; 
dotted line: $T^*/\omega_E = 0.11 \,\sqrt{\lambda-1.49}$ for $\mu^*=0.24$. 
The inset shows the determinant in (\ref{eq:det})
as a function of $T$ for $\mu^*=0$ and 
$\lambda = 0.3$, $0.4$, $0.5$, $0.7$ and $1$ (from top to bottom).}
\end{figure}

\section{Discussion and conclusion}
\label{sec:5}

Our method provides a novel way to calculate $T_c$ of
superconductors, starting from the normal state.
At the superconducting instability, Eliashberg's equations are derived 
from the RG
flow equations. These self-consistent integral
equations (\ref{matsubara_1}) are 
equivalent to the condition of zero determinant
expressed in (\ref{eq:det}). In our RG approach, once the bare values of the
couplings are given, we simply calculate the flow of 
$\det[1+{\bf U}(0) \cdot {\bf P}(T)]$ by solving the differential
RG flow equations and determine $T^*$ as the point when this quantity becomes
zero (see inset in Fig.\ref{numerics}). This numerical calculation is much
simpler than solving (\ref{matsubara_1}) directly, since it does
not involve self-consistency conditions.  
In Fig.\ref{numerics} we show the calculation of $T^*$ as a
function of $\lambda$ with this new method together with the
approximate asymptotic expressions for weak and strong coupling.

In summary, we have developed an asymptotically exact RG method
that takes into account electron-electron and electron-phonon
interactions in an unbiased way and reproduces all of Eliashberg's
theory and also provides a framework (large-$N$) for
understanding it as well as Migdal's theorem that goes into it.
Our procedure can be used for any Fermi surface geometry and for
any number of scattering channels (forward, charge and spin
density wave, etc) and therefore allows for the study of the
competition between scattering channels \cite{next}.  Finally, our
procedure allows for a new numerical way to investigate
superconductivity in metals.

\acknowledgments
We thank J.~Carbotte, A.~Chubukov, C.~Chamon, J.~B.~Marston,
G.~Murthy, and M.~Silva-Neto for illuminating discussions
and the Aspen Center for Physics for its hospitality during the
early stages of this work. A.~H.~C.~N. was supported
by NSF grant DMR-0343790. R.~S. was supported by NSF grant
DMR-0103639.


\begin{thebibliography}{20}
\bibitem{screening}For simplicity we assume screened electron-electron
interactions although the screening can be also derived \cite{shankar}.
\end{thebibliography}
\end{document}